\documentclass[aps,prb,reprint,amsmath,amssymb,floatfix]{revtex4-2}

\usepackage{graphicx}
\usepackage{bm}
\usepackage{siunitx}
\usepackage[colorlinks=true,allcolors=blue]{hyperref}

\graphicspath{{./}{./defect_figures/}}

\begin{document}

\title{Defect Candidates for Single-Photon Emission in Monolayer GaS and GaSe}

\author{Michael Lorke}
\email{michael.lorke@uni-due.de}

\affiliation{Faculty of Physics, University of Duisburg-Essen, 47057 Duisburg, Germany}

\begin{abstract}
We investigate the spin-resolved electronic structure of native vacancies and substitutional defects in monolayer GaS and GaSe using an anisotropic screened-exchange functional developed for two-dimensional material systems. By combining an energy-based classification of defect-associated states with spatially resolved projections and partial charge densities, we examine how these defects modify the in-gap electronic structure and assess their relevance for defect-based single-photon emission. 
The results identify gallium vacancies, particularly in GaS, as candidates single-photon emitters.
\end{abstract}

\maketitle

Point defects provide a means of engineering the electronic structure of two-dimensional (2D) semiconductors at the atomic scale. Vacancies and substitutional impurities can alter the local coordination, change the occupation of frontier orbitals, and introduce states within the host band gap. Depending on their charge state, symmetry, and coupling to lattice vibrations, such states can contribute to carrier trapping, magnetism, and optoelectronic response \cite{Hong2017}. Defects are also of interest as possible solid-state single-photon emitters (SPEs) \cite{Toth2019}. 

Gallium monochalcogenides GaS and GaSe are layered III--VI semiconductors whose electronic and optical properties are sensitive to thickness, structural environment, and chemical modifications \cite{Ma2013}. Their electronic structure exhibits features that are not adequately described by a simple parabolic band-edge picture. In particular, monolayer GaSe exhibits a pronounced Mexican-hat-shaped valence band around $\Gamma$ \cite{BenAziza2018,huang2016anisotropic}, while the optical response of monolayer GaS is sensitive to the electronic structure and symmetry of the states involved \cite{Guo2017}. These characteristics make it important to distinguish genuine defect-induced levels from states that merely acquire appreciable weight near a defect while remaining part of the host-band manifold.

Previous first-principles studies have examined native defects in these materials. Intrinsic defects in monolayer GaS were investigated using semilocal density-functional calculations, supplemented by selected comparisons with the HSE06 functional, revealing defect-induced states and magnetic behavior \cite{Chen2015}. Subsequently, native defects in monolayer GaS and GaSe were studied with respect to their electronic properties and thermodynamic stability using semilocal and screened-hybrid approaches \cite{BarraganYani2022}. These studies provide a basis for understanding the electronic effects of native defects, but the potential of defect configurations as SPEs raises additional questions concerning their in-gap electronic structure and possible optical transitions.

Recent theoretical investigations have proposed selected point defects in group-III monochalcogenides as potential SPEs \cite{Stolbov:2022}. Related calculations have also explored local defects in monolayer GaS as potential SPE candidates \cite{Stolbov:2023}. These predictions motivate further investigation of defect-associated electronic states in both materials. 

The present work focusses on cation substitutional configurations and intrinsic defects in monolayer GaS and GaSe using a consistent electronic-structure approach. We examine the spin-resolved defect-associated states and identify configurations whose electronic structures may warrant further investigation as potential SPE candidates. The exchange-correlation treatment is important for this comparison because semilocal and screened approximations can suffer from self-interaction and delocalization errors. We therefore employ a recently developed anisotropic screened-exchange functional for two-dimensional material systems \cite{Lorke:26}. The functional is designed to account for anisotropic screening in two-dimensional materials and to restore piecewise-linear behavior of the total energy with fractional occupation, thereby reducing self-interaction and delocalization errors. To ensure a consistent comparison, we apply the same functional to the structural relaxation and subsequent electronic-structure calculations. 

We investigate Si and Zn substitutions on Ga sites, Ga vacancies, and chalcogen vacancies in monolayer GaS and GaSe. 
By combining the electronic structure of the defective systems with spatially resolved characterisation of defect-associated states, we establish a comparative picture of how these defects modify the in-gap electronic structure and identify transitions that may be relevant to defect-based SPEs.


The dataset contains eight defect configurations, four in each host. The spin-resolved calculations contain 480 bands and use a 6x6x1 supercell in the \(\Gamma\)-approximation. The pristine-host valence-band maximum (VBM) and conduction-band minimum (CBM) 
at this same supercell \(\Gamma\) point 
define the common energy reference for classifying defect states in both spin channels for each material. 

The calculations were performed within density-functional theory using VASP and the projector augmented-wave method \cite{Blochl:94,Kresse:99}. Defective monolayers were modeled with \(6\times6\times1\) supercells; the vacancy configurations contain 143 atoms. A vacuum region of \(\SI{20}{\angstrom}\) separated periodic images normal to the layer. Dipole corrections were applied as implemented in VASP to reduce spurious electrostatic interactions between repeated slabs. Long-range dispersion was included using the Tkatchenko--Scheffler correction \cite{Tkatchenko:09}. The structures and electronic states were calculated with the anisotropic screened-exchange functional  \cite{Lorke:26}; this functional was used for both relaxation and final electronic-structure calculations of the pristine and defective systems. The plane-wave kinetic-energy cutoff was \(\SI{450}{\electronvolt}\), and the Ga PAW datasets included Ga \(d\) states in the valence. Structures were relaxed until residual Hellmann--Feynman forces were below \(\SI{1e-3}{\electronvolt\per\angstrom}\). All defect calculations were performed for the neutral charge state. Collinear spin was used, and the tabulated \(\mu\) values are the final total magnetic moments of the defective supercells.


In GaS, the relaxed defect geometries exhibit distinct changes in the local coordination. For $\mathrm{Si}_{\mathrm{Ga}}$, the three nearest S atoms are approximately $2.200$~\AA{} from Si, compared with a pristine Ga--S distance of $2.367$~\AA{}, corresponding to a reduction of about $7.1\%$. Substitution by Zn instead gives nearest-neighbour distances of approximately $2.322$~\AA{}, only $1.9\%$ shorter than the pristine Ga--S distance. For the Ga vacancy, the three nearest S atoms lie approximately $2.221$~\AA{} from the nominal vacancy site, while the three nearest Ga atoms around the S vacancy are approximately $2.156$~\AA{} from its reconstructed position. Both vacancy structures therefore exhibit a contraction of the nearest-neighbour shell relative to the corresponding pristine coordination distance. These differences show that the local structural response depends on the defect species and whether the defect occupies a Ga or S site.

In GaSe, the pristine Ga--Se nearest-neighbour distance is approximately $2.437$~\AA{}. For $\mathrm{Si}_{\mathrm{Ga}}$, the three nearest Se atoms lie approximately $2.362$~\AA{} from Si, a reduction of about $3.1\%$. The corresponding distances for $\mathrm{Zn}_{\mathrm{Ga}}$ are approximately $2.439$--$2.442$~\AA{}, essentially unchanged from the pristine distance. Around the Ga vacancy, the three nearest Se atoms are approximately $2.283$--$2.289$~\AA{} from the reconstructed vacancy position, whereas the three nearest Ga atoms around the Se vacancy are approximately $2.141$--$2.145$~\AA{} from its reconstructed position. Thus, the substitutional defects differ markedly in their local structural response, while both vacancy structures show a contracted nearest-neighbour shell. The small spread in the vacancy-neighbour distances also indicates that the relaxed local environments are not perfectly equivalent in all directions.

\begin{figure}[t]
\includegraphics[width=\columnwidth]{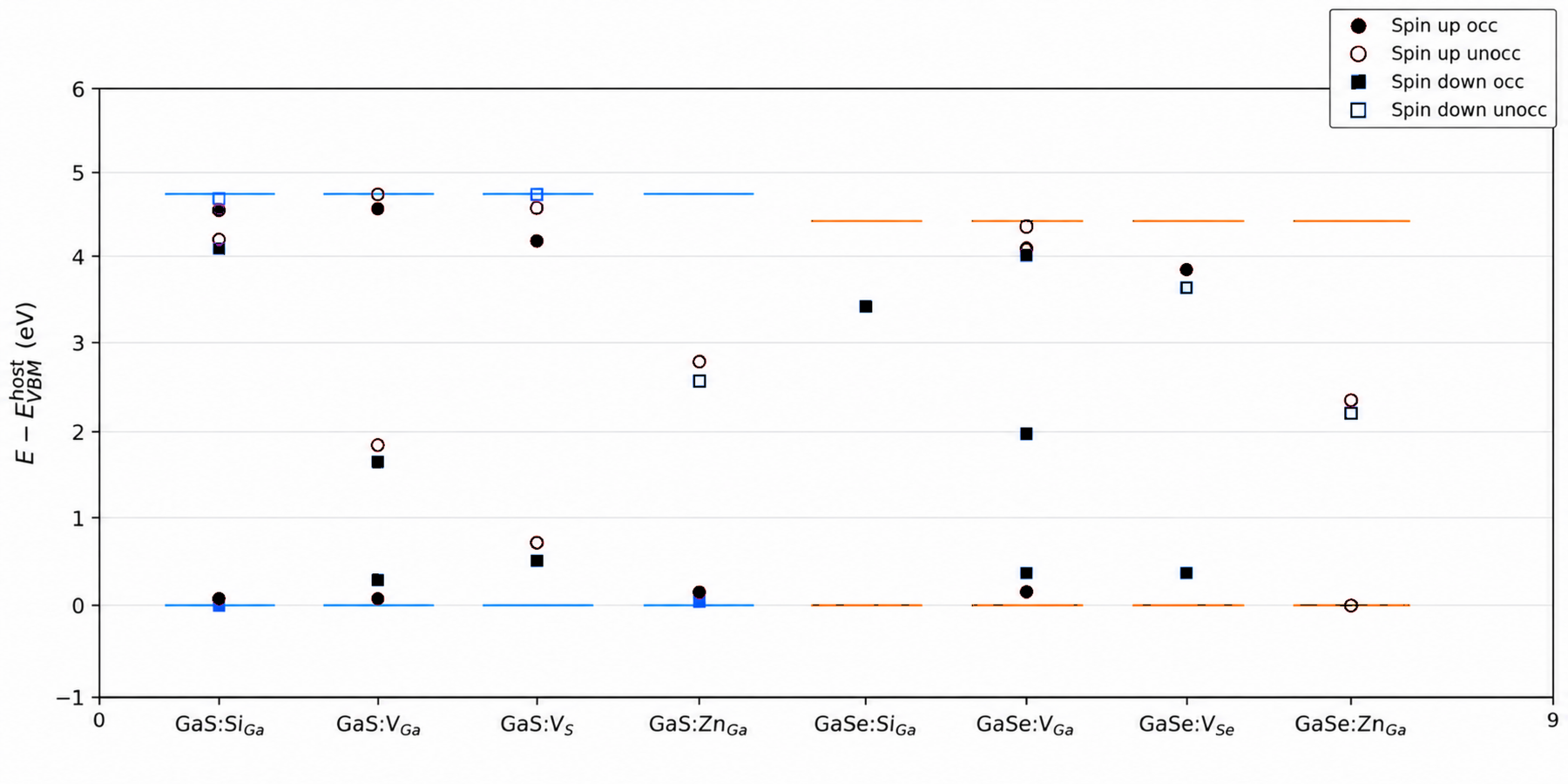}
\caption{\label{fig:levels} Defect states of the eight relaxed-geometry defect calculations, aligned to the pristine-host $\Gamma$-point VBM. Filled (open) symbols denote (un-)occupied states. Circles and squares distinguish the two spin channels. The host edges are those of the pristine cell.}
\end{figure}

In the electronic structure calculations, we compare the defect calculation with the pristine-host VBM and CBM. For states within the bands a weight criterion $w_n>0.10$ is used to screen for appreciable defect-region character. 
Figure~\ref{fig:levels} presents the resulting classification for all localized levels in the eight calculations. Energies are referenced to the pristine-host VBM and CBM.

\begin{table}[t]
\caption{\label{tab:electronic} Lowest spin-resolved Kohn--Sham transitions of the defective supercells. The magnetic moment $\mu$ is the final total moment of the full defective supercell.}
\begin{ruledtabular}
\begin{tabular}{lccc}
System & $\mu$ ($\mu_B$) & $\Delta E^\uparrow$ (eV) & $\Delta E^\downarrow$ (eV)\\
\hline
GaS:Si$_{\mathrm{Ga}}$   & $-1$ & 4.650 & 0.573\\
GaS:V$_{\mathrm{Ga}}$    & $+1$ & 4.308 & 1.472\\
GaS:V$_{\mathrm{S}}$     & 0    & 3.465 & 3.465\\
GaS:Zn$_{\mathrm{Ga}}$   & $+1$ & 4.620 & 2.721\\
GaSe:Si$_{\mathrm{Ga}}$  & $-1$ & 4.488 & 0.963\\
GaSe:V$_{\mathrm{Ga}}$   & $+1$ & 3.787 & 1.496\\
GaSe:V$_{\mathrm{Se}}$   & 0    & 3.385 & 3.385\\
GaSe:Zn$_{\mathrm{Ga}}$  & $+1$ & 4.494 & 2.294\\
\end{tabular}
\end{ruledtabular}
\end{table}

We find a marked difference in the energetic alignment of defect-associated states across the investigated configurations. Most notably, GaS:\(V_{\mathrm S}\) and GaS:\(\mathrm{Zn}_{\mathrm{Ga}}\) exhibit both occupied and unoccupied states within the host band gap, making them promising candidates for further investigation of defect-related optical transitions. For GaS:\(V_{\mathrm S}\), the occupied states lie close to the valence-band maximum (VBM), while the unoccupied states occur near the conduction-band minimum (CBM). GaS:\(\mathrm{Zn}_{\mathrm{Ga}}\) likewise exhibits in-gap states, with a pronounced spin dependence in their energetic alignment.

\begin{figure}[t]
\includegraphics[width=\columnwidth]{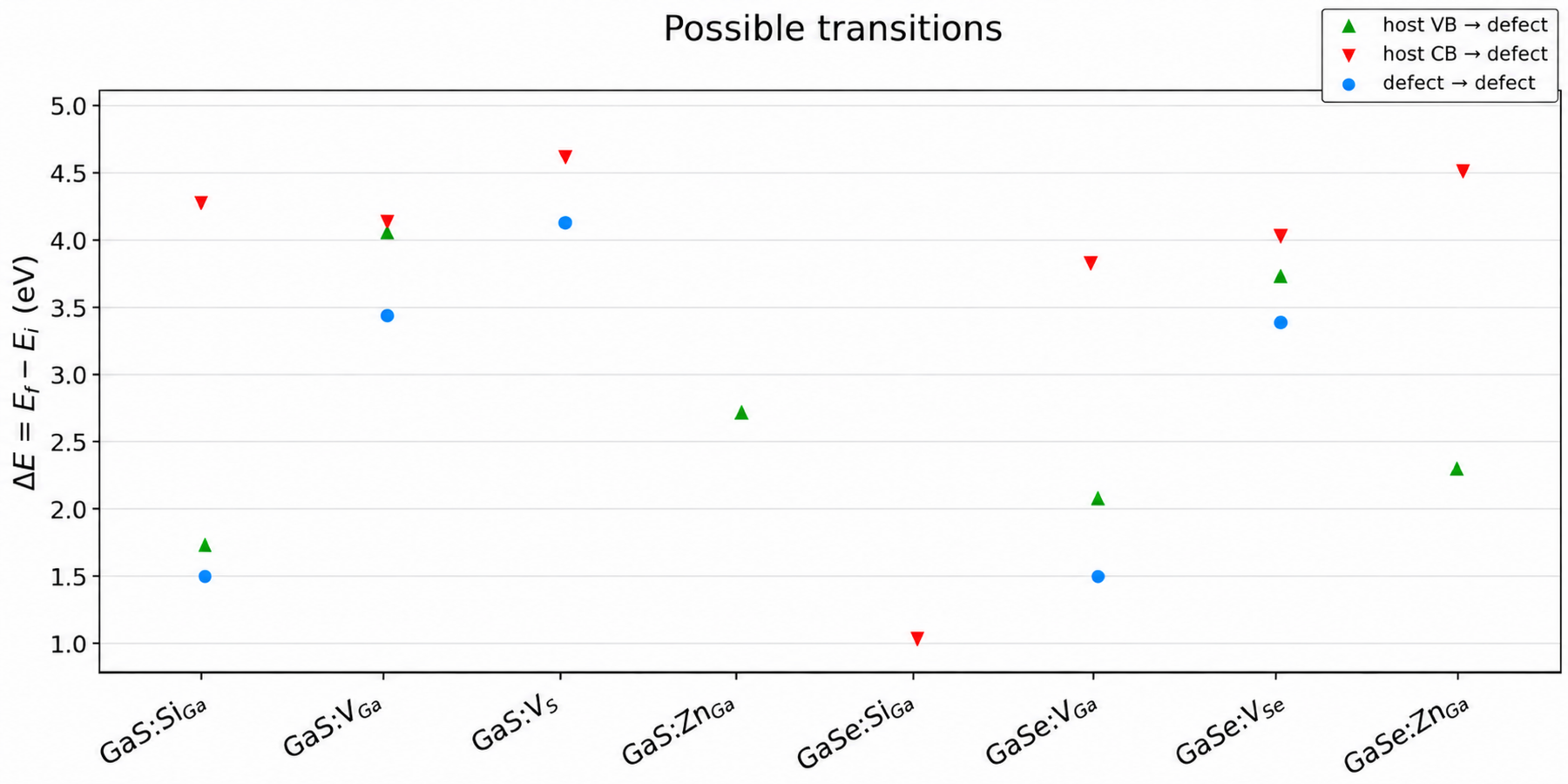}
\caption{\label{fig:transitions}Spin-diagonal Kohn--Sham transitions for the relaxed geometries. }
\end{figure}

Figure~\ref{fig:transitions} summarizes possible transitions by configuration and  class. 
The gallium vacancies in GaS and GaSe are particularly noteworthy, with defect-to-defect Kohn–Sham energy differences of 1.472 and approximately 1.496 eV, respectively. By comparison, the corresponding defect-to-defect transitions for the chalcogen vacancies occur at higher energies, 3.465 eV for GaS:\(V_{\mathrm S}\) and approximately 3.385 eV for GaSe:\(V_{\mathrm{Se}}\).

\begin{table}[t]
\caption{\label{tab:transitions}
Lowest-energy spin-conserving Kohn--Sham transition candidates
in each category. Energies are given in eV.}
\begin{ruledtabular}
\begin{tabular}{lccc}
System &
VB $\to$ defect &
CB $\to$ defect &
Defect $\to$ defect \\
\hline
GaS:Si$_{\mathrm{Ga}}$  & 1.73  & 4.28  & --- \\
GaS:V$_{\mathrm{Ga}}$   & 4.05  & 4.13  & 1.472 \\
GaS:V$_{\mathrm{S}}$    & ---   & 4.63  & 3.465 \\
GaS:Zn$_{\mathrm{Ga}}$  & 2.72  & ---   & ---  \\
GaSe:Si$_{\mathrm{Ga}}$ & ---   & 1.04  & ---  \\
GaSe:V$_{\mathrm{Ga}}$  & 2.08  & 3.83  & 1.496 \\
GaSe:V$_{\mathrm{Se}}$  & 3.73  & 4.03  & 3.385 \\
GaSe:Zn$_{\mathrm{Ga}}$ & 2.30  & 4.52  & ---  \\
\end{tabular}
\end{ruledtabular}
\end{table}

To verify the optical activity, we examined the independent-particle dielectric response
of the defective cell. As an example GaS:V$_{\mathrm{Ga}}$ is shown, where the in-plane dielectric response exhibits a low-energy feature at approximately $1.5$ eV, which can be attributed to the transition with a Kohn–Sham eigenvalue difference of 1.472 eV. 

\begin{figure}[t]
\centering
\includegraphics[width=\columnwidth]{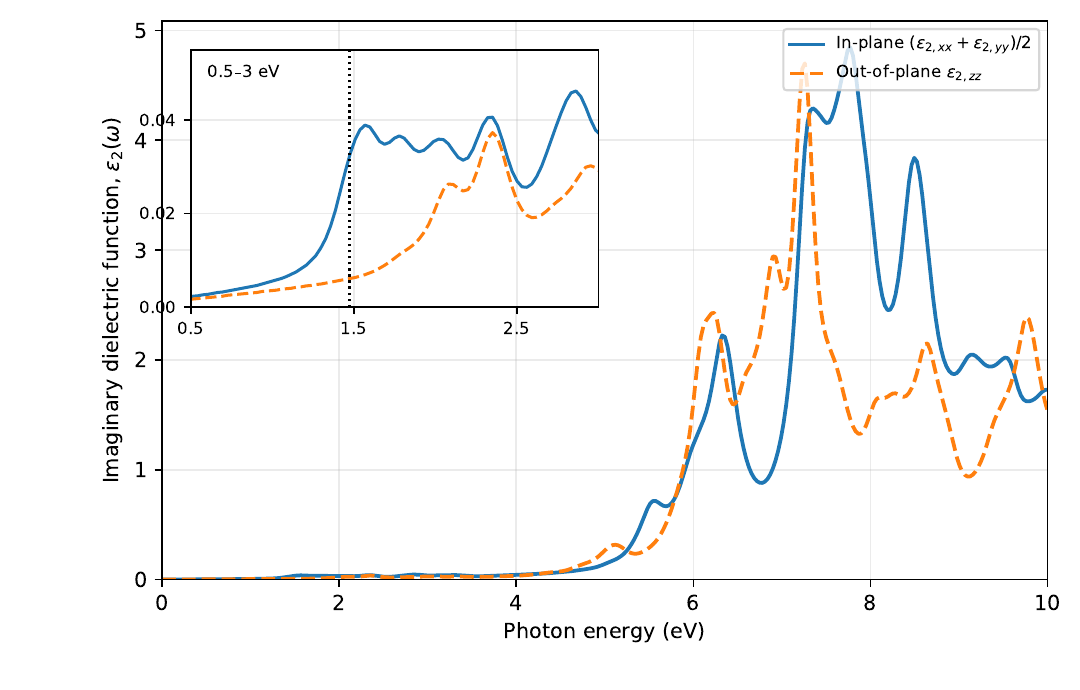}
\caption{
Imaginary part of the dielectric tensor for GaS${\mathrm{Ga}}$, showing the in-plane average $(\varepsilon{2,xx}+\varepsilon_{2,yy})/2$ and the out-of-plane component $\varepsilon_{2,zz}$. The inset enlarges the low-energy region.}
\label{fig:optical}
\end{figure}

\begin{figure*}[t]
\centering
\begin{minipage}[t]{0.245\textwidth}
\centering
\includegraphics[width=\linewidth]{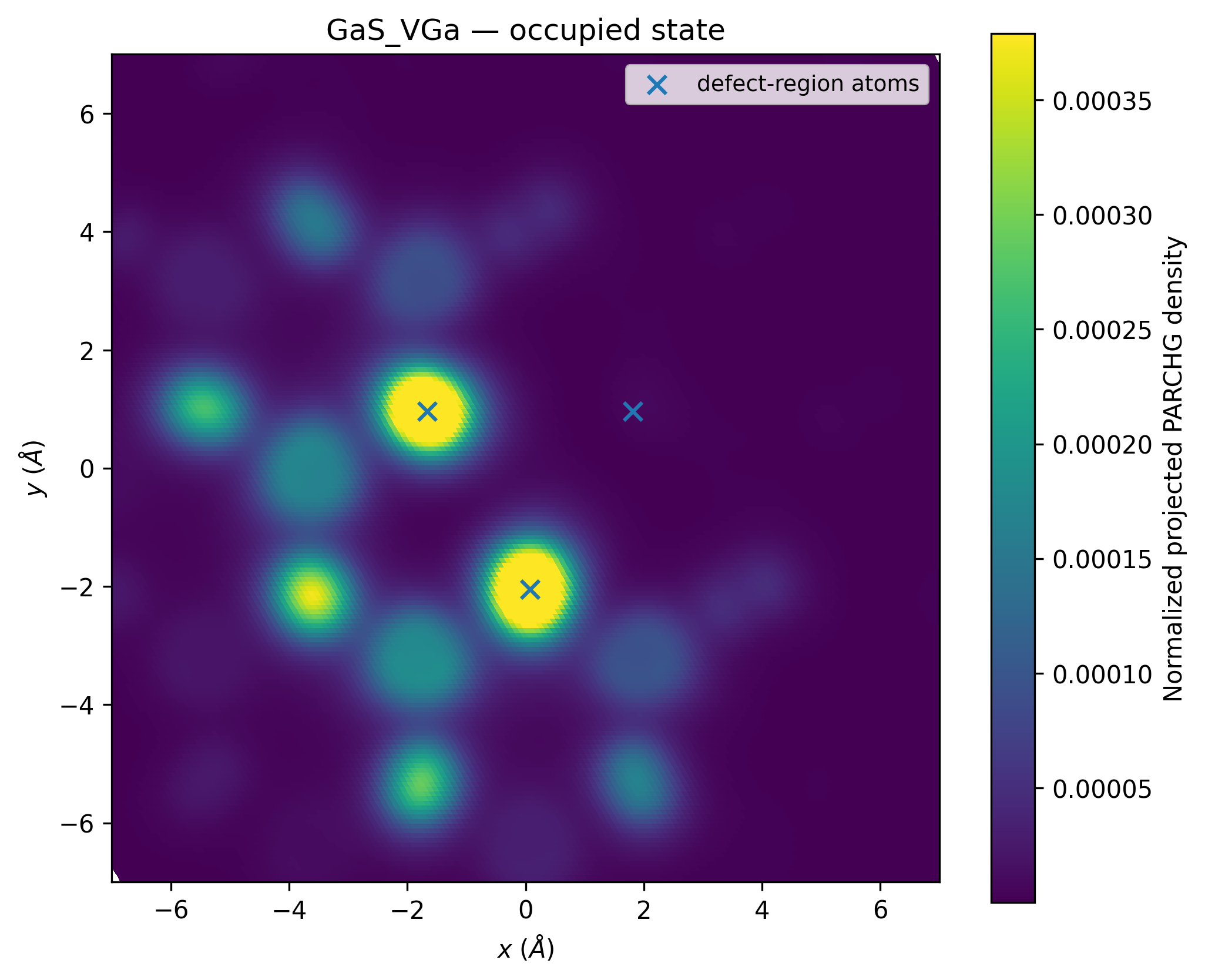}
\small (a) GaS:V$_{\mathrm{Ga}}$, occupied
\end{minipage}
\begin{minipage}[t]{0.245\textwidth}
\centering
\includegraphics[width=\linewidth]{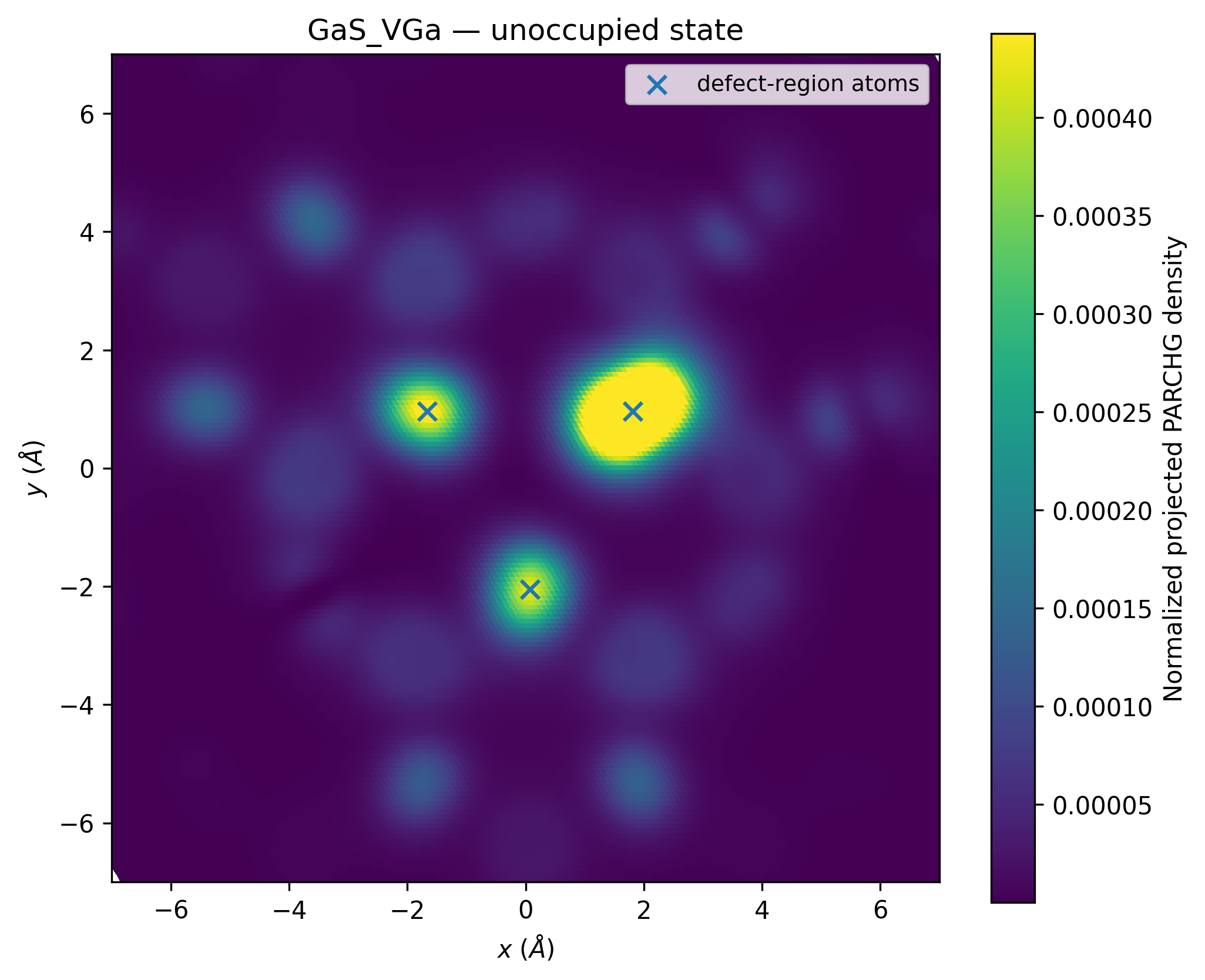}
\small (b) GaS:V$_{\mathrm{Ga}}$, unoccupied
\end{minipage}
\begin{minipage}[t]{0.245\textwidth}
\centering
\includegraphics[width=\linewidth]{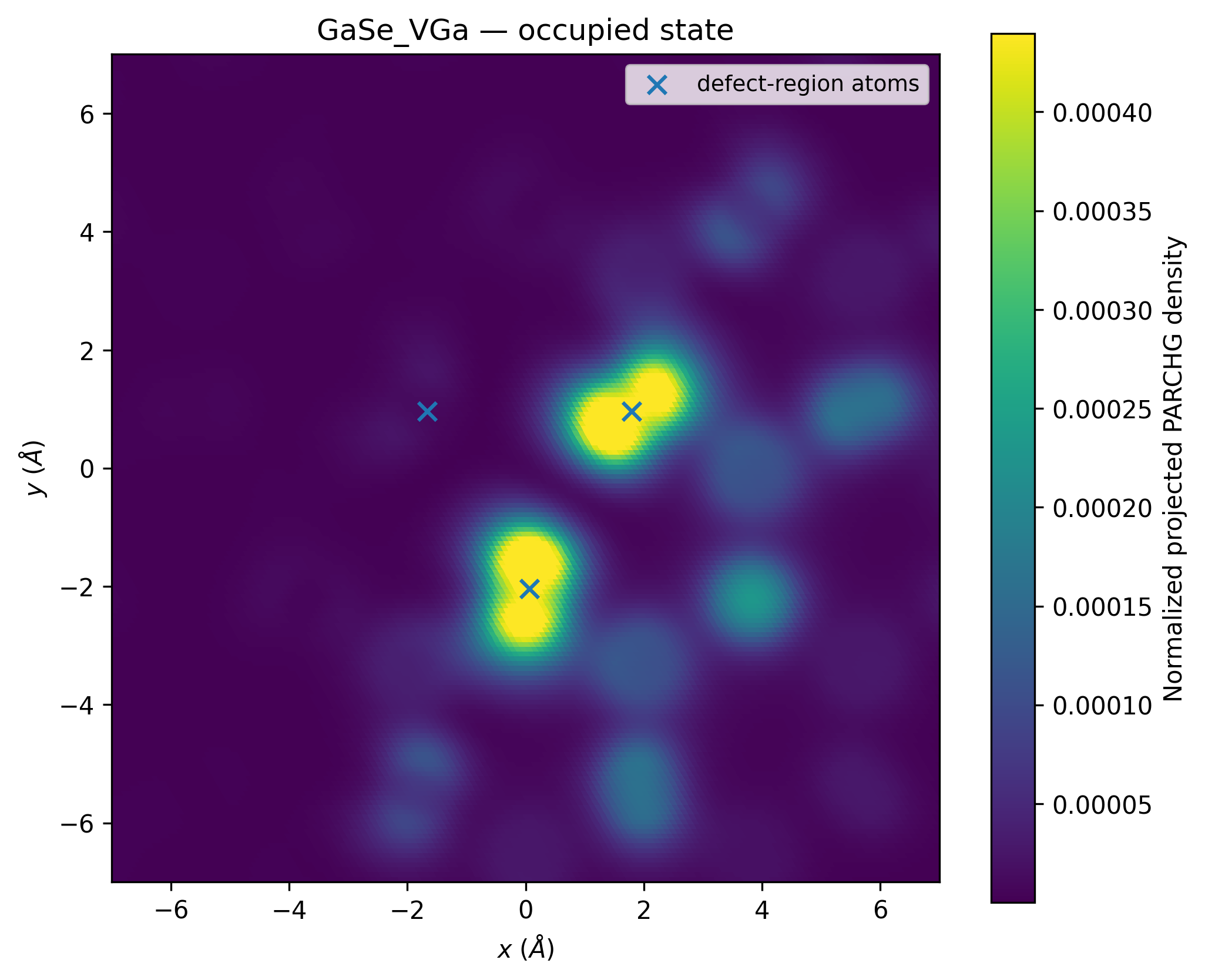}
\small (c) GaSe:V$_{\mathrm{Ga}}$, occupied
\end{minipage}
\begin{minipage}[t]{0.245\textwidth}
\centering
\includegraphics[width=\linewidth]{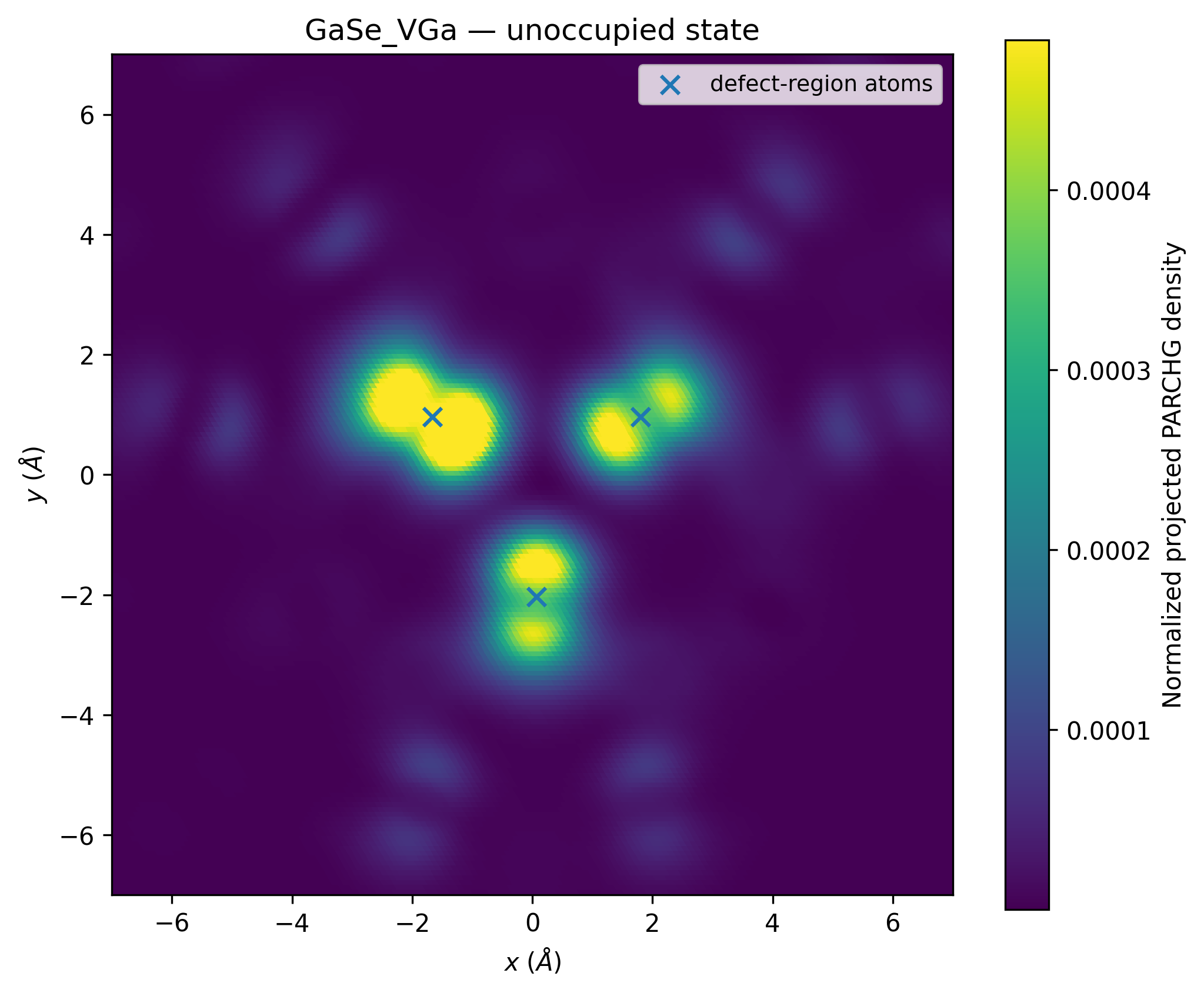}
\small (d) GaSe:V$_{\mathrm{Ga}}$, unoccupied
\end{minipage}
\caption{\label{fig:ga-vacancies}In-plane projected partial charge densities for selected occupied and unoccupied states associated with gallium vacancies. The two materials show similar qualitative contrast between a more extended occupied state and a more compact unoccupied state, particularly for GaSe:V$_{\mathrm{Ga}}$. Each panel uses its own displayed density scale, so color intensity should not be compared quantitatively between panels. }
\end{figure*}

\begin{figure*}[t]
\centering
\begin{minipage}[t]{0.245\textwidth}
\centering
\includegraphics[width=\linewidth]{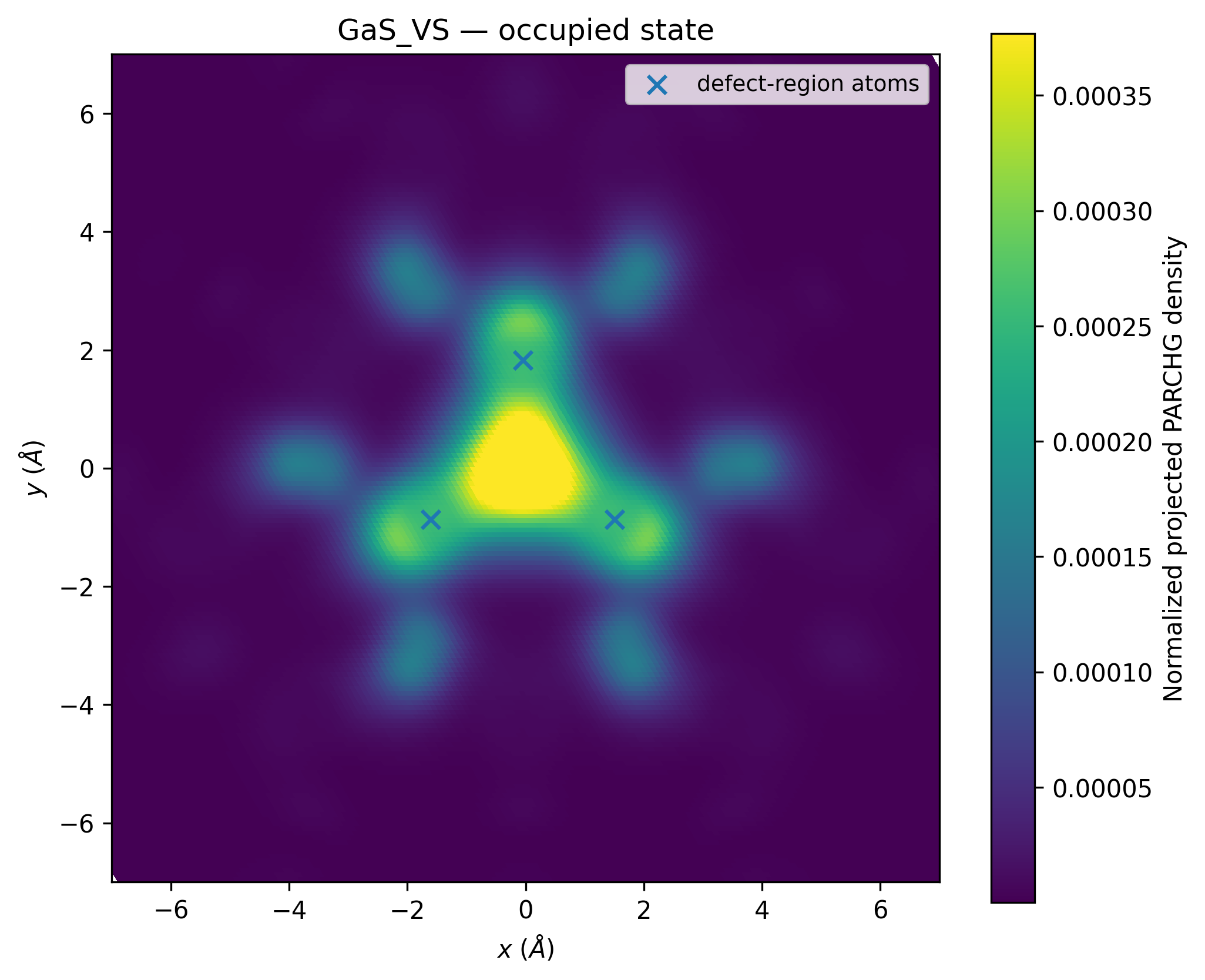}
\small (a) GaS:V$_{\mathrm{S}}$, occupied
\end{minipage}
\begin{minipage}[t]{0.245\textwidth}
\centering
\includegraphics[width=\linewidth]{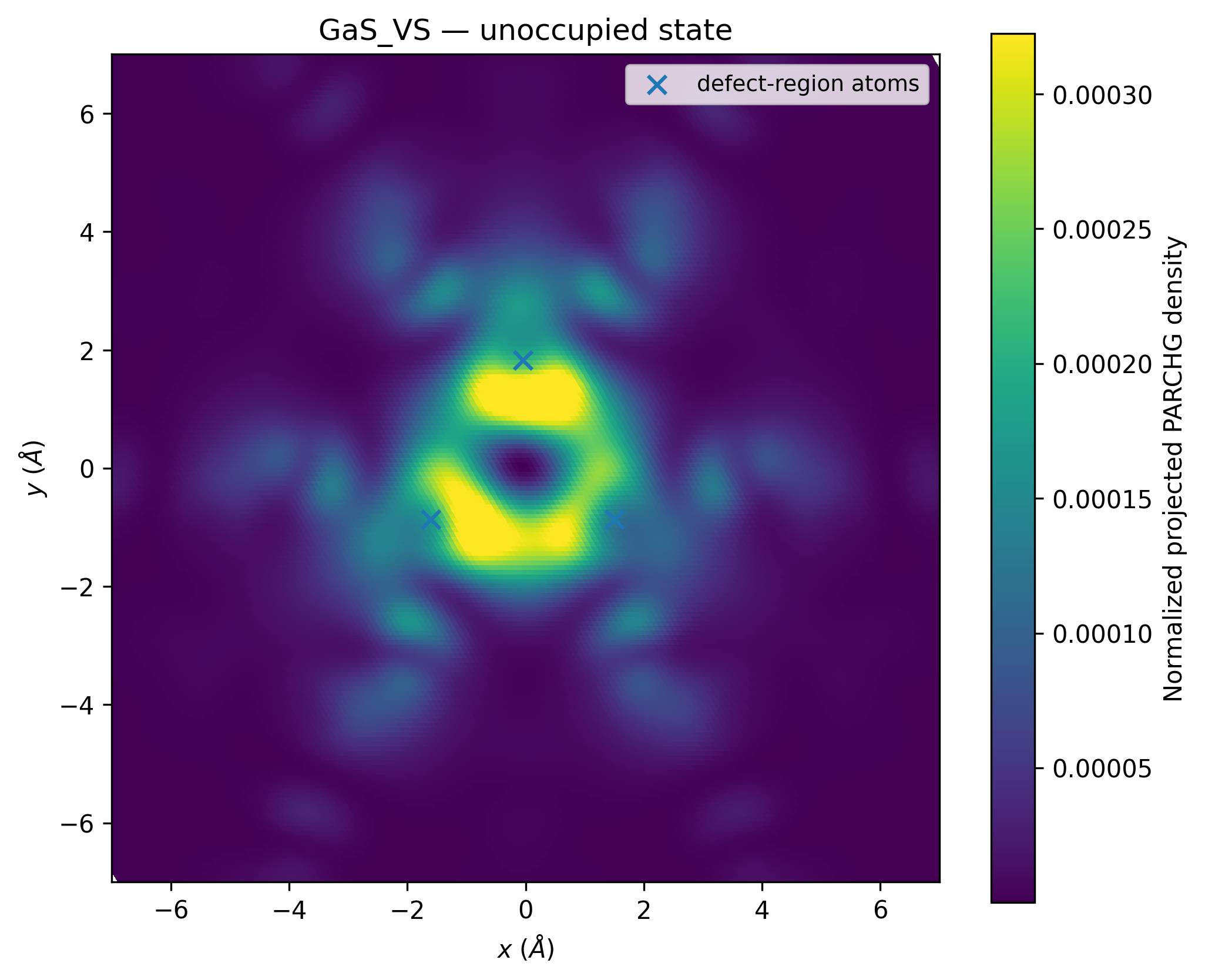}
\small (b) GaS:V$_{\mathrm{S}}$, unoccupied
\end{minipage}
\begin{minipage}[t]{0.245\textwidth}
\centering
\includegraphics[width=\linewidth]{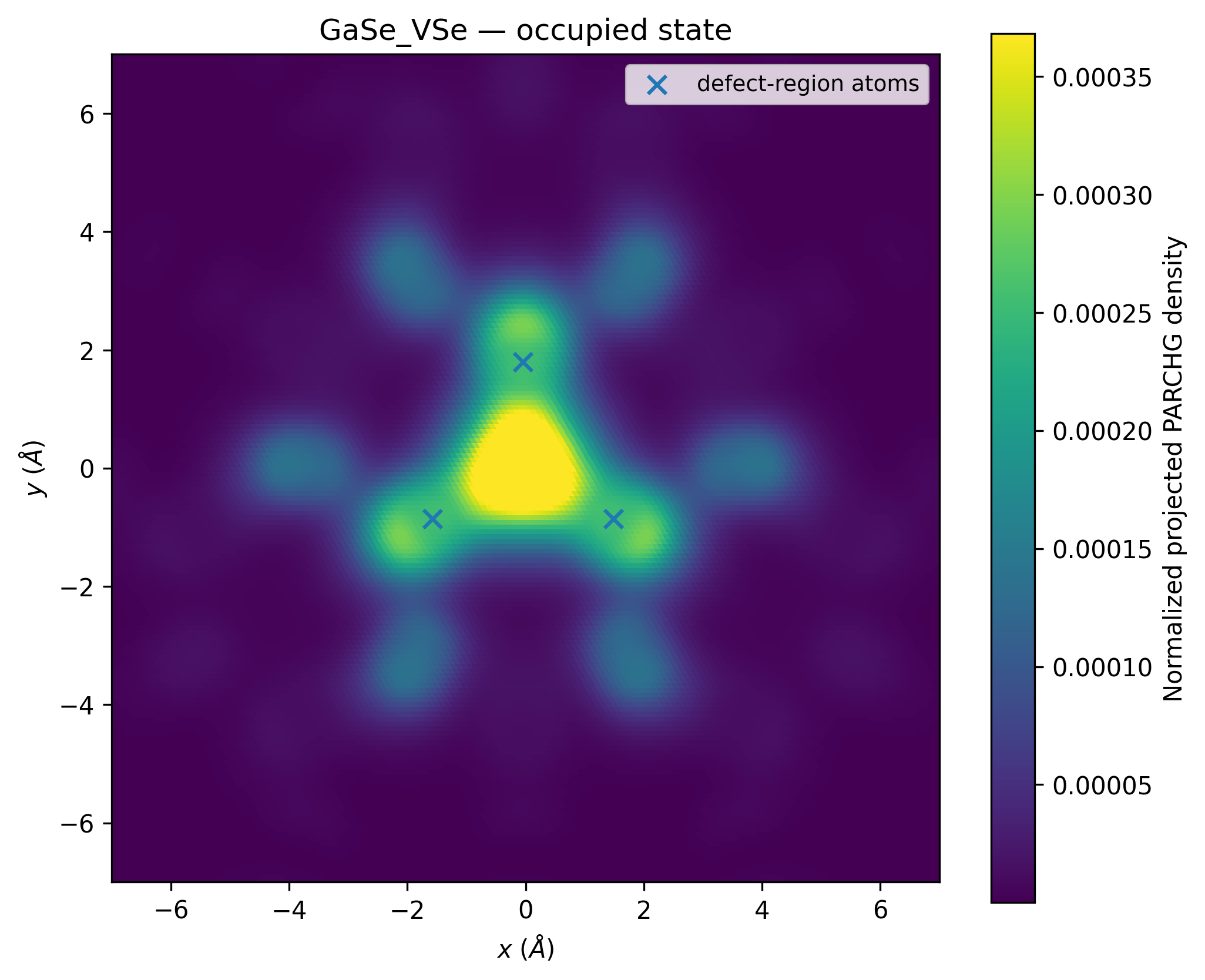}
\small (c) GaSe:V$_{\mathrm{Se}}$, unoccupied, band 322
\end{minipage}
\begin{minipage}[t]{0.245\textwidth}
\centering
\includegraphics[width=\linewidth]{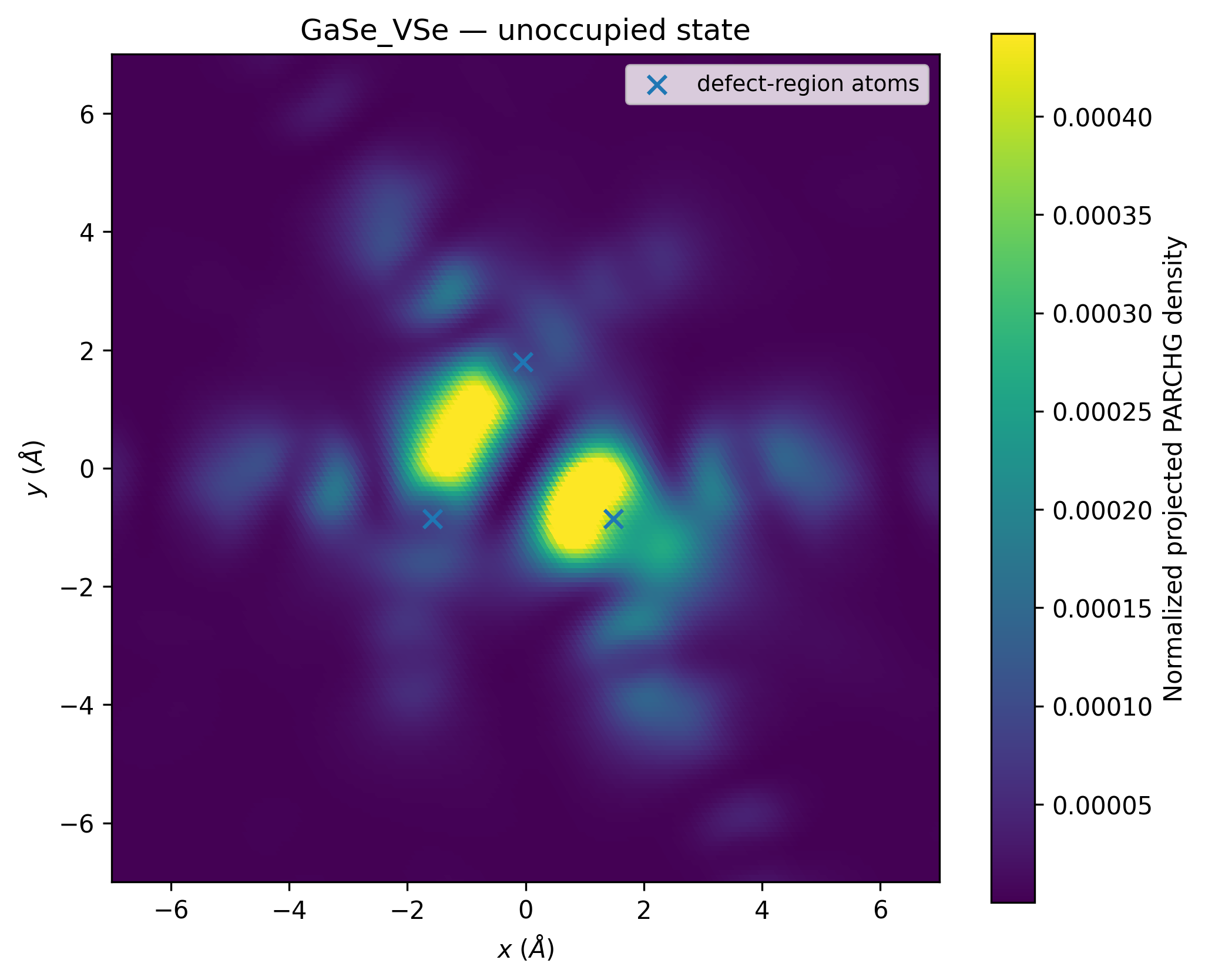}
\small (d) GaSe:V$_{\mathrm{Se}}$, unoccupied, band 323
\end{minipage}
\caption{\label{fig:x-vacancies}In-plane projections of individual Kohn--Sham partial charge densities for selected chalcogen-vacancy states. Panels (c) and (d) show separate PARCHG files for GaSe:V$_{\mathrm{Se}}$ bands 322 and 323, respectively; they are not summed over bands. Each panel is independently scaled, so color intensity is not a cross-panel measure of localization.}
\end{figure*}

Figures~\ref{fig:ga-vacancies} and \ref{fig:x-vacancies} compare selected occupied and unoccupied states for the gallium and chalcogen vacancies. The displayed maps are in-plane projections of the partial charge density; crosses mark exactly the atoms included in the defect-region projection sum $w_n$ defined in Sec.~II.

The maps provide qualitative evidence that the selected vacancy-associated states have nonuniform charge density around the defect region. They do not, on their own, quantify a localization length or establish a dipole-allowed optical transition. Because each map is displayed on its own color scale, the color intensity cannot be used to rank localization across panels. The two unoccupied GaSe:V$_{\mathrm{Se}}$ maps correspond to distinct Kohn--Sham bands 322 and 323 and are shown separately; neither panel represents a sum over those bands.

We have investigated the spin-resolved electronic structure of native vacancies and substitutional defects in monolayer GaS and GaSe using an anisotropic screened-exchange functional for two-dimensional materials. The comparison reveals substantial differences in the energetic alignment, spin dependence, and transition energies of the defect-associated states, providing a basis for prioritizing configurations for further optical investigation.

Among the configurations considered, the Ga vacancy in GaS, \(V_{\mathrm{Ga}}\), emerges as the most promising candidate. Its occupied and unoccupied defect-associated states exhibit a Kohn–Sham energy separation of 1.472 eV. The corresponding Ga vacancy in GaSe is a second promising candidate, with a defect-to-defect energy separation of approximately 1.496 eV.
The chalcogen vacancies in GaS and GaSe also exhibit defect-associated states, but their lowest defect-to-defect energy separations are much higher, at 3.465 and 3.385 eV, respectively, deep in the ultra-violet. 
The substitutional configurations provide additional possibilities, particularly \(\mathrm{Zn}_{\mathrm{Ga}}\), which exhibits occupied and unoccupied in-gap states in GaS. These comparisons demonstrate that candidate selection depends on the combined electronic structure, spatial character of the participating states, and optical response.
Overall, the results identify \(V_{\mathrm{Ga}}\) in GaS as the leading candidate for subsequent defect-specific optical calculations, with \(V_{\mathrm{Ga}}\) in GaSe as a closely related system for comparison. 

M.L acknowledges funding by the Deutsche Forschungsgemeinschaft (DFG, German Research Foundation) under grant number LO 1840/7-1 as well as support from the DFG via SFB 1242 (Project-ID 278162697) and CPU time at HLRN (Berlin/G\"ottingen).

\section*{Data Availability Statement}
The data supporting the findings of this study are available from the corresponding author upon reasonable request.

\bibliographystyle{apsrev4-2}
\bibliography{defect_screening_refs}

\end{document}